\documentclass[aps,prd,preprint,superscriptaddress,showpacs,showkeys]{revtex4}
\usepackage{amssymb,amsmath}
\usepackage{graphicx} 
\usepackage{dcolumn}  
\usepackage{epsfig}   
\usepackage{pstricks}
\usepackage{ifpdf}
 \usepackage[greek, english]{babel}
\begin{document}


\title{ Neither a field nor a matter: The case of energy}


%
%

\author{D-M. Cabaret}
\affiliation{UMR-CNRS 8167, membre associ\'e}
\email[]{dominiquemarie.cabaret@gmail.com}

\author{T. Grandou}
\email[]{thierry.grandou@orange.fr}

\author{G.-M. Grange}
\affiliation{Couvent des Dominicains\\
1, Impasse Lacordaire\\
Toulouse\\
31400 France}
\email[]{gm.grange@gmail.com}

\author{E. Perrier}
\affiliation{ISTA\\
1, Impasse Lacordaire\\
Toulouse\\
31400 France}
\email[]{perrier@revuethomiste.fr}


\date{\today}

\begin{abstract} 
Energy is no doubt an intuitive concept. Following a previous analysis on the nature of elementary particles and associated elementary quantum fields, the peculiar status and role of energy is scrutinised further at elementary and larger scales. Energy’s physical characterisation shows that it is a primordial component of reality highlighting the quantum fields’ natural tendencies to interact, the elementary particles’ natural tendency to constitute complex bodies and every material thing’s natural tendency to actualise and be active. Energy therefore is a primordial notion in need of a proper assessment.
\end{abstract}


\maketitle

\section{\label{SEC:1}Introduction}
One is so accustomed to the daily use of the word `energy', and the various forms of experience one can make of it, that, clearly, the issue of knowing what it is indeed, is barely considered a relevant questioning. Yet, if one tries to go beyond this first draft evidence and investigates the matter a bit further, then this apparent evidence is discovered to recede endlessly. B. Pascal once noticed \cite{Pascal}, \emph{"La fin des choses et leurs principes sont pour lui [l'homme] invinciblement cachés dans un secret impénétrable."} ${}^{\footnotemark[1]}$. \footnotetext[1]{For him [Man], the end of things and their principles are invincibly hidden in an impenetrable secret.} This secrecy seems to have continued until today, since, as can be learned out of the Web, \emph{Nobody knows what the actual energy is} \cite{Web}.
\par
Aristotle coined the term \emph{energeia} (\emph{\textgreek{ἐνέργεια}})) \cite{Aristotle} to account for the activity of something. As such, it was opposed to the \emph{dunamis} (\emph{\textgreek{δύναμις}}), the power, or capacity, or ability, which is the origin (external or internal) of this thing’s activity. Subsequently, energeia and dunamis
could be used to qualify the state a reality is in: A thing is in-act or in-potency. It is in-act in
regard to the achievement brought about by the activity, and it is in-potency in regard to
what is not yet actualised. Therefore, the becoming of realities through movement or change can be described as the transition from one state to another state, from being-in-potency to being-in-act. For example, the acorn is a potential oak tree (state of being-in-potency) while the adult tree is an actual oak tree (state of being-in-act). 
\par
After this philosophical creation, the term energy was reused in the 19th century in thermodynamics,
in association with the notion of work (Thomas Young, 1802), that is, what is done with
energy. In other words, energy was then construed from the point of view of the activity’s
physical effect. Thus, the principle of the conservation of energy was posed and the
notions of kinetic energy and potential energy made it possible to describe the movements
of mobiles (an historical account can be found in \cite{Bruce}). The
notion of energy is now essential in physics, but again, it is difficult to grasp the reality it
covers.

\par
Nowadays, scientists define energy as the ability to do work. Modern civilisation is possible because people have learned how to change energy from one form to another and then use it to do work. People use energy to walk and bicycle, to move cars along roads and boats through water, to cook food on stoves, to make ice in freezers, to light our homes and offices, to manufacture products, and to send astronauts into space. That is, several different forms of energy can be noticed, such as light, heat, motion, the electrical, chemical and gravitational forms, and a convenient classification for the purpose of doing work relies on the two sorts of energy's forms, the potential or stored energy and the kinetic or working energy forms \cite{web}.
\par
Now, the concept of energy remains \emph{very subtle..}, as R.P. Feynman once quoted, \emph{.. it is very difficult to get it right} \cite{Feynman}.
\par\medskip
The goal of the current paper is precisely to strive to grasp a better and deeper insight on energy, on what it is in itself. The paper is organised as follows. 
\par
In Section II, energy is scrutinised at the elementary scale. For this, the elements of the material world are recalled in some details. Emphasis is put on the fact that these elements cannot be envisaged independently of their fundamental interactions because they \emph{are} and \emph{are in order to..}: Though there is room to consider elementary field elements in themselves, it is notoriously incomplete and thus highly misleading to conceive them as if they were inert and static things to be stared at in a museum. The hallmark of the elementary scale is argued to be the famous Einstein $E=mc^2$ relation, while the possibility of a `pure energy' is denied, so as for energy the possibility that it be a `thing'. 
\par
In particular, in complement to a previous analysis \cite{A3}, new arguments are proposed to discard an identification of energy to the {\emph{primary matter}}, long sought  by philosophers and physicists. A final subsection entails five points to be retained out of this long Section.
\par
In the third Section the relation of mass and energy is analysed at scales larger than the elementary, and the autonomy of mass with respect to energy is discussed by taking advantage of the \emph{covariant} form of the Einstein relation. The translation of the physical considerations into a well defined metaphysical understanding of the subject reveals to be straightforward and precious in order to understand and to order the facets that energy displays all over the various scales of the physical world.
\par 
The fourth section finally addresses the central question of what energy is. From the physical knowledge of the material world, in effect, several operative definitions of energy have been derived which are equivalent at the dimensional level, but do not deliver any information concerning the very nature of energy. Out of quantum field theories, which scrutinise matter at the elementary scale, a clear-cut characterisation of energy has finally emerged, while recognising the notion of energy as a primary one. The help of philosophical arguments will reveal to be efficient in order to clarify the content of the energy's notion, and to reach two \emph{indirect} definitions of it.
\par
Finally, a fifth Section attempts to summarise things as simply as possible within a language rid of the too many technical terms, which are otherwise in order in the main text of the paper. Note that at several steps, this analysis relies on results which have been derived in the recent analysis of Reference \cite{A3}, whereof its frequent mentions.
\section{At the elementary scale}
In a previous and recent publication the very first elements of our material World were identified to be the elementary quantum fields associated to the known, reported elementary particles. The latter being a measurable and highly specified mode of actualisation of the former which, themselves, are not measurable entities \cite{A3}. Letting {\emph{gravitation}} aside, the other three forces which shape our Universe are the electromagnetic force, the weak and the strong forces, all three of them being most efficiently described by the so-called {\emph{Standard Model of elementary particles}}. In these three cases, forces are mediated by exchanges of {\emph{bosonic}} fields' quanta, the photon $\gamma$, the intermediate bosons, $Z^0, W^\pm$, and the gluons $G$. Wether gravitation proceeds in the same way, exchanging would-be {\emph{gravitons}} is not known at present, and in any case, an overall unification of all four forces is not yet available.

\subsection{The elements of the physical world}
In \cite{A3}, the very first determinations of our material World were identified as the 17 elementary quantum fields, some of them are explicitly recalled below for the sake of illustration, the photonic quantum field to begin with (here written in the so-called {\emph{Coulomb gauge}}),
\begin{equation}\label{photon}
A_\mu(t,\vec{x})=\sum_{s=1,2}\int {d^3k\over (2\pi)^3\,2\omega(\vec{k})}\,\biggl[\varepsilon_\mu^{(s)}(\vec{k})a_s(\vec{k})e^{-ik_0 t+i\vec{k}\cdot\vec{x}}+{\varepsilon_\mu^{(s)}}^*(\vec{k})a^\dagger_s(\vec{k})e^{+ik_0 t-i\vec{k}\cdot\vec{x}}\biggr]
\end{equation}
Similar expressions can be written down explicitly also for the intermediate \emph{bosonic} fields the $Z^0$, the $W^\pm$ and the eight gluonic fields $A^a_\mu$. 
For the electronic field, \emph{fermionic}, one has,
 \begin{equation}\label{fermionic1}
\Psi(t,\vec{x})=\sum_{s=1,2}\int {d^3p\over (2\pi)^3\,{\sqrt{2E_p}}}\,\biggl[ u^s(p)b_s(\vec{p})e^{-ip_0 t+i\vec{p}\cdot\vec{x}}+{v^s}(\vec{p})d_s^\dagger(\vec{p})e^{+ip_0 t-i\vec{p}\cdot\vec{x}}\biggr]
\end{equation}and its conjugate field associated to the \emph{positron}, the {\emph{anti-particle}} of the electron,
 \begin{equation}\label{fermionic2}
{\bar{\Psi}}(t,\vec{x})=\sum_{s=1,2}\int {d^3p\over (2\pi)^3\,{\sqrt{2E_p}}}\,\biggl[ {\bar{v}}^s(p)d_s(\vec{p})e^{-ip_0 t+i\vec{p}\cdot\vec{x}}+{{\bar{u}}^s}(\vec{p})b_s^\dagger(\vec{p})e^{+ip_0 t-i\vec{p}\cdot\vec{x}}\biggr]\,,
\end{equation}with, encoded in their quantisation algebras, the fundamental repulsive character of fermions. For (\ref{fermionic1}) and (\ref{fermionic2}) one has, 
\begin{equation}\label{quant1}
\{d_s(\vec{p}),d^\dagger_{s'}(\vec{p'})\}=\{b_s(\vec{p}),b^\dagger_{s'}(\vec{p'})\}= (2\pi)^3\delta _{ss'}\delta^{(3)}(\vec{p}-\vec{p'})\,.\end{equation}the other possible anti-commutators being zero, while the tendency of bosons to agglutinate is also encoded in their respective quantisation rule,
\begin{equation}\label{quant2}
[a_s(\vec{p}),a^\dagger_{s'}(\vec{p'})]= (2\pi)^3\delta _{ss'}\delta^{(3)}(\vec{p}-\vec{p'})\,,\end{equation}the other possible commutators being zero also ${}^{\footnotemark[2]}$. \footnotetext[2]{Commutators are defined by $[a,b]=ab-ba$ and anti-commutators by $\{a,b\}=ab+ba$.}
\par
  Of course, in themselves, these expressions are mathematical entities only, technically known as \emph{operator valued distributions on some relevant representation spaces}; and for this reason, they should rather be referred to as `quantised fields'. Now, as argued in \cite{A2} for the wave function of Quantum Mechanics, these quantised fields {\emph{do}} refer to the genuine physical quantum realities, which will be referred to as `quantum fields', that quantised fields describe so accurately for a large span of physical situations for which they have been conceived; not for all situations though \cite{efimov}.
In \cite{A3}, these quantised fields have been dubbed \emph{elements} in reference to the \emph{antic} cosmological concepts of that time.
\par
From the point of view of physics, now, the 17 elements-fields no longer have much to do with the elements as they were conceived in antiquity. But from the point of view of the \emph{philosophy of nature}, they occupy the same place and play the same function of first degree of material realities, and subsequently, as the first determinations of the \emph{primary matter}. 
\par\medskip
Quantum fields, in other words, would therefore stand for {\textit{the very first physical realities composing in the physical bodies of our material Universe \cite{A3}}}. 
\par\medskip
Now, there is more to it. This is essentially because the Universe is not a giant pool table, made out of balls that some external forces would move into some random game in order to produce, in the long term, the highly structured Universe which we know. The balls in effect, that is the elementary particles associated to the quantum fields (\ref{photon}) or (\ref{fermionic1}) \emph{are}, and \emph{are in order to}: They are in order to compose into the more substantial realities of the hadronic matter, and itself, to compose into atom's nuclei at a later stage.
\par
This means that the elementary particles are not to be stared at as if they were like motionless statues in a museum, since their very nature opens them to the fundamental interactions they are ordered to, and thanks to which, it is worth recalling here, they could precisely be discovered.
\par
Accordingly, it seems appropriate to think of the elementary building blocks giving rise to the material World as given by the elementary quantum fields {\emph{together with}} their fundamental interactions. Along this view, elementary quantum fields such as (\ref{photon})-(\ref{fermionic2}) get completed into the full building blocks of the Quantum Electro Dynamical world (the $QED$ theory) by identifying the fundamental electron-positron-photon vertex of interaction ${}^{\footnotemark[3]}$ \footnotetext[3]{In gauge theories, interactions may be described in ways which differ from (\ref{emint}), the so-called minimal coupling. Now, the quantum fields themselves may be redefined \cite{Zee}, in particular through the unavoidable renormalisation algorithm. The point is that all of these technical steps preserve the observable/measurable quantities the quantum theories give access to, and that whatever their forms in various choices of gauges, interactions remain.},
\begin{equation}\label{emint}
-ie\gamma_\mu\,{\bar{\Psi}}(t,\vec{x})A^\mu(t,\vec{x})\Psi(t,\vec{x})\end{equation}
where the $\gamma_\mu$ stand for the four Dirac matrices, $\mu\in \{0,1,2,3\}$ and $+e$ for the electric charge of the positron. A diagrammatic representation of (\ref{emint}) can be given as in Fig.1 and is but a simple example of the well-known \emph{Feynman diagrams} so widely used in \emph{perturbative treatments} of quantum field theories.
\begin{figure}[!htb]
   \includegraphics[scale=0.3]{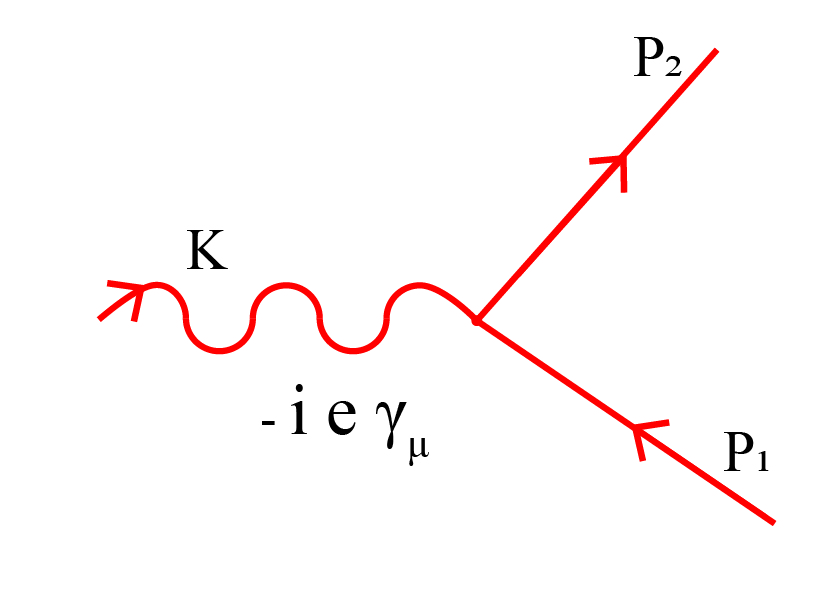}
   \caption{At the electromagnetic interaction vertex, with coupling $-ie\gamma_\mu$, an incoming electron with 4-momentum $P_1$ absorbs a photon of 4-momentum $K$ and is scattered off with 4-momentum $P_2=P_1+K$.}
   \label{fig1}
\end{figure}

\par
 Another example, which will be referred to later on (see Fig.4) can be given with the quark-antiquark-gluon interaction vertex pictured on Fig.2.
 \begin{figure}[!htb]
   \includegraphics[scale=0.3]{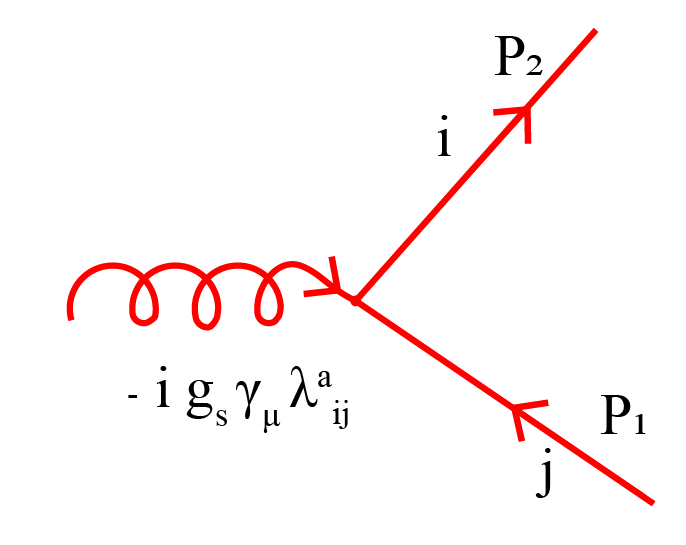}
   \caption{At the strong interaction vertex, with coupling $-ig_s\gamma_\mu\lambda^a_{ij}$, an incoming quark with 4-momentum $P_1$ and color index $j$ absorbs a gluon of 4-momentum $K$ and is scattered off with 4-momentum $P_2=P_1+K$ and color index $i$.}
   \label{fig 2}
\end{figure} 
 Up to the additional complexities inherent to the $SU_c(3)$ non-abelian gauge group which is relevant to the the strong interactions, this vertex is analogous to the previous one (\ref{emint}), and reads ${}^{\footnotemark[4]}$, \footnotetext[4]{Repeated indices, $a$ and $\mu$, are summed upon and $f$ runs over the full set of the 6 \emph{flavours} of quark fields, {\textit{u,d,s,c,b,t}}. The $\lambda^a$ stand for the 8 generators of the $SU(3)$-Lie algebra taken, here, in the so-called \emph{fundamental representation} where quark fields `live', $a=1,2,\dots,8$.}
\begin{equation}\label{qbarqg}
-ig_s\sum_f\gamma^\mu\,{\bar{q_f}}^j(t,\vec{x})A_a^\mu(t,\vec{x})\lambda_{ij}^a\,q_f^i(t,\vec{x})\end{equation}where $g_s$ is the \emph{strong coupling constant} and $\lambda_{ij}^a$ the $ij$ element of the $SU(3)$-Lie algebra matrix $\lambda^a$. In a similar way, one can give the interaction's vertices involving the standard model bosons $W^+$ and $W^-$ and $Z^0$ in their interactions with quarks and leptons; these are textbook material \cite{PS}. However, writing them all in full extent would take pages. For our current concern, fortunately, it is enough to give a few details. The full interaction reads as~\cite{PS},
    \begin{equation}\label{SM}
\mathcal{L}_{int.}=g\,(W^+_\mu J^{\mu +}_W+W^-_\mu J^{\mu -}_W+Z^0_\mu J^\mu_{Z})+e\,A_\mu J^\mu_{EM}
\end{equation}where, in particular,
\begin{equation}\label{term1}
J^{\mu +}_W=\frac{1}{\sqrt{2}}(\bar{\nu}_L\gamma^\mu e_L+{\bar{u}}_L\gamma^\mu d_L)\,, \ \ \ \ \ \ J^{\mu -}_W=\frac{1}{\sqrt{2}}(\bar{e}_L\gamma^\mu \nu_L+{\bar{d}}_L\gamma^\mu u_L)\end{equation}
and the last term of (\ref{SM}),
\begin{equation}\label{em}
J^\mu_{EM}=\bar{e}\gamma^\mu(-1)e+\bar{u}\gamma^\mu(+\frac{2}{3})u+\bar{d}\gamma^\mu(-\frac{1}{3})d\end{equation} is the standard electromagnetic current, coupling to the photonic field $A_\mu$. Not necessary to our current considerations, the involved expression of $J^\mu_Z$ \cite{PS} can be omitted here.
\par
In (\ref{term1}), one observes the electronic field, denoted by $e$ and its associated neutrino, $\nu_e$. The subscript $L$ accounts for the \emph{Left handed} character of neutrino fields \cite{PS}, while quark fields themselves can be written as a sum of Left and Right handed components : The former only couple to the $W^\pm_\mu$ bosons. 
\par
In (\ref{em}), one can observe the presence of the quark fields $u$ and $d$ with, in units of $+e$ ({\textit{i.e.}}, the positron charge), their respective electric charges of $+\frac{2}{3}$ and $-\frac{1}{3}$. It is interesting also to notice the flavour-changing character of the two first interaction terms in (\ref{SM}).
\par
These $u$ and $d$ quarks are those of the so-called \emph{first generation} composed of these two quarks, and of the two leptons, the electron $e$ and its neutrino $\nu_e$, the 1st generation, $(u,d,e,\nu_e)$. Two extra generations of increasing masses exhaust the Standard Model fermionic content : The 2nd generation comprising the quarks $s$ and $c$, the lepton $\mu$ and its neutrino $\nu_\mu$, that is $(s,c,\mu,\nu_\mu)$; and the 3rd generation, $(b,t,\tau,\nu_\tau)$, comprising the quarks $b$ and $t$, the lepton $\tau$ and its neutrino $\nu_\tau$. The very existence of these $3$ generations is still quite enigmatic an issue in particle physics ${}^{\footnotemark[5]}$. \footnotetext[5]{In spite of a few interesting insights, provided for instance by some Clifford algebraic considerations \cite{Daviau}, these $3$ generations are essentially taken to be a primary data of Nature.}
 Quark fields therefore interact with the photonic field $A_\mu$, as well as with the electroweak fields $W^\pm$ and $Z^0$, while their proper strong interactions, described by the theory of \emph{Quantum Chromo Dynamics} ($QCD$), are mediated instead by the $8$ gluonic fields, the $A^a_\mu$-fields of (\ref{qbarqg}), which interact also among themselves.
 \par
It is remarkable that the visible matter of the Universe comes from the first generation, while the particles of the second and third generations  are unstable and quickly decay into first generation particles which are stable. \par\medskip
To sum up, the elementary \emph{building blocks} of the material Universe can now be conceived in view of the last experimental developments of physics, and they correspond to the 17 (or 18, if the graviton is discovered) elementary quantised fields associated to the registered elementary particles. Through their interactions with each others, these first \emph{elements} compose into the more elaborate bound states which give rise progressively to the ordinary matter. 
\par
This is why the Universe elementary building blocks are to be conceived as the elementary quantised fields \emph{together with} their interaction diagrams : The physical realities corresponding to the elementary quantised fields \emph{are}, and \emph{are in order to}, as is manifest both at the actual and purely potential levels of the physical reality \cite{A3}.
\par\medskip
It is worth emphasising again that the above paragraphs can be taken from the attestation that the interaction diagrams derive from fields' properties : They belong to the very nature of the quantum fields and actual interactions proceed according to the diagrams encrypted in the fields' nature. This has consequences for taking us away from an understanding of elementary particles such as inherited from atomism.
Interactions of elementary particles are not accidental and should not be conceived outside of the quantum fields' realities and detached from their definitions. 
\par
In this subsection, thus, the elements of the physical world have been recalled in order to manifest the omnipresence of energy at the very first step of the corporeal realities, and also to stress an aspect of paramount importance, which is that of a thorough conservation of energy and momentum within each interaction process. This stands for a universal law, satisfied at any level of the physical reality, actual and potential, like vacuum fluctuations in this latter case.

\subsection{Energy at the elementary scale, or $E=mc^2$}
At the elementary scale, which is that of elementary particles and of their associated quantum fields, it is a most remarkable fact that $99\%$ of the hadronic matter mass is obtained out of \emph{massless} fields \cite{Wilczek}. This is a non-trivial statement if we keep in mind that `Newtonian mechanics posited mass as a primary quality of matter, incapable of further elucidation. We now see Newtonian mass as an emergent property' \cite{Wilczek}:
\par

\emph{Most of the mass of standard matter, by far (as much as $99\%$!), arises dynamically, from back-reaction of the color gluon fields of quantum chromodynamics (QCD). Additional quantitatively small, though physically crucial, contributions come from the intrinsic masses of elementary quanta (electrons and quarks), themselves related to the Higgs mass, as well as the $W^\pm$ and $Z$ masses \cite{Wilczek}.}
\par
This checks the universal and well known equivalence of mass and energy, so concisely expressed by the most famous Einstein equation $E=mc^2$,${}^{\footnotemark[6]}$\footnotetext[6]{This most famous equation is not relativistically invariant, contrarily to the equation (\ref{invariant}) given below. This means that $E=mc^2$ is valid in a particular reference frame only (see p.21).} made even more explicit in the \emph{natural unit system} where, in particular, $c=1$. And effectively, elementary particles' masses are given in units of energy, in $MeV$, or $GeV$ for the heavier ones (where the $eV$, the \emph{electron-volt}, is but a matter of metrological convention, fitted to the energy scale being considered).
\par\medskip
The elementary scale, that is the material world at the elementary scale, is thus the realm whose hallmark sign could be taken to be the famous $E=mc^2$ relation. This is also why the special theory of relativity has been perceived so perfectly fitted to the description of the microphysical world \cite{Piersaux}, and proven to be so \cite{kerner}, though not {\textit{ab initio}} conceived and derived in this very context.
\par
Within the theories of relativistic quantised fields, the terms resulting from a perturbative expansion in the interaction coupling strengths (the constants $e$, $g_s$ and $g$ of (\ref{emint}), (\ref{qbarqg}) and (\ref{SM}) respectively)   are commonly represented by Feynman diagrams, as alluded to above. These diagrams may or may not directly correspond to physical processes ${}^{\footnotemark[7]}$\footnotetext[7]{They don't in general and in particular in the Standard Model of particle physics whose fundamental principle, dynamical, is that of \emph{local gauge invariance.}}. In either cases, however, they involve the equivalence of mass and energy, and their mutual conversion. 
 \begin{figure}[!htb]
   \includegraphics[scale=0.3]{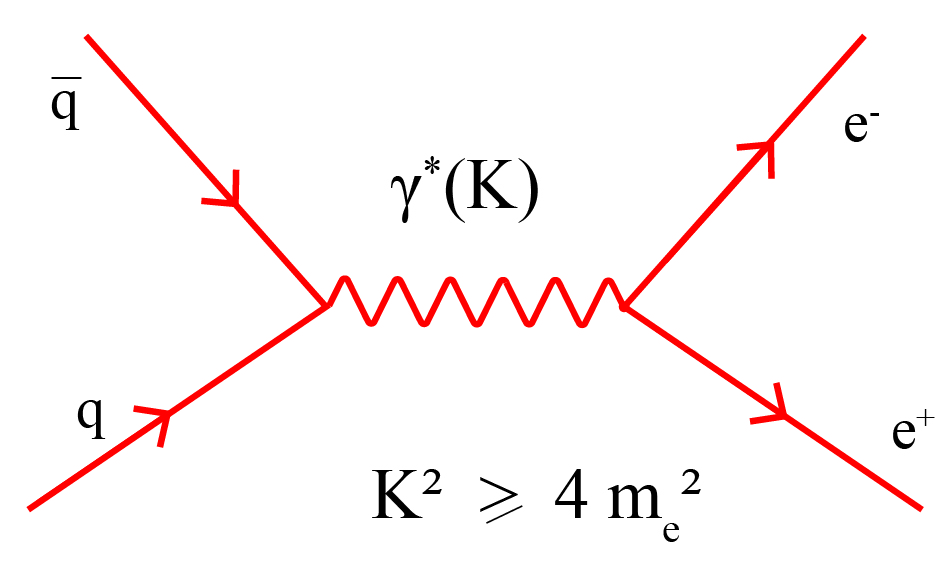}
   \caption{A quark and an anti-quark annihilate into a virtual photon $\gamma^\star(K)$ which decays into an electron/positron pair, provided that the relation $K^2\geq 4m_e^2$ is satisfied. Four-momenta conservation is checked at each vertex of interaction.}
   \label{fig 3}
\end{figure} 
On Fig.3, for example, an energetic enough excitation in the photonic quantum field, a $\gamma^\star(K)$, is pictured to produce a real electron/positron  pair, $e^+,e^-$. It is noteworthy that the same $\gamma^\star(K)$ can be experimentally realised by crossing intense laser beams \cite{HMF}. Conversely, the excited photonic line which appears in the middle of Fig.3 can be generated by the annihilation of a matter-anti-matter pair, a pair of quark-anti-quark for example (Fig.3 left hand side), and in this very case one deals with the so-called \emph{Drell-Yan mechanism of lepton pair production} {\cite{T-M. Yan}, thoroughly described within perturbative $QCD$, and beyond, by the standard model ${}^{\footnotemark[8]}$\footnotetext[8]{This process was used to design the experiments at CERN that discovered the W and Z bosons and was crucial in the discovery of the \emph{top quark} at Fermilab. The discovery of the Higgs boson at CERN in 2012 is perhaps the most dramatic example of the utility of the Drell-Yan mechanism \cite{T-M. Yan}.}. There exists a number of other similar examples connecting particles which are not \emph{directly} related to each others by the fundamental interactions, but which can nevertheless interact, mainly through photonic excitations, and others also, like $Z$-excitations in the present Drell-Yan case. This exhibits a somewhat \emph{transversal} aspect of the photonic field, an example of which is pictured on Fig.4.
 \begin{figure}[!htb]
   \includegraphics[scale=0.3]{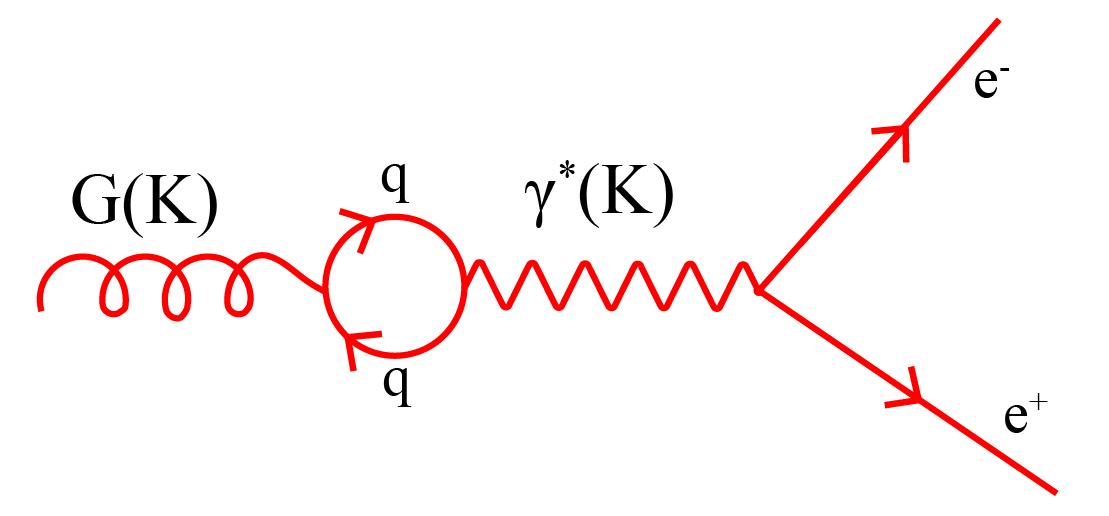}
   \caption{A virtual gluon $G$, with 4-momentum $K$ such that $K^2\geq 4m_e^2$, fluctuates into a quark loop which connects to a virtual photon $\gamma^\star(K)$, ultimately decaying into an electron/positron pair. Four-momenta conservation is checked at each vertex of interaction.}
   \label{fig 4}
\end{figure} 
\subsubsection{A `pure energy' ?}
This noticeable aspect of the photonic field excitations, mediated by so-called \emph{virtual photons} (ordinarily denoted by the symbol $\gamma^\star(k)$ in the scientific literature), shows up in this {{transversal}} way among the whole span of elementary particle interactions and has led physicists to think of it as `pure energy' itself. To wit,
\par
\emph{The energy is not a thing, not a property as well as not a condition. The energy is the existence of the state of the mass. The energy is made up of radiation only. Nobody knows what the actual energy is \cite{Web}.}
\par
And also,
\par
\emph{Energy is in fact the
substance from which all elementary particles, all atoms and therefore all things are made,
and energy is that which moves. Energy is a substance, since its total amount does not
change, and the elementary particles can actually be made from this substance as is seen
in many experiments on the creation of elementary particles. Energy can be changed into
motion, into heat, into light and into tension. Energy may be called the fundamental cause
for all change in the world \cite{WH}.}
\par\medskip
As will be recalled in next subsection C, it is worth noting that this latter statement of W. Heisenberg, amounts to an implicit identification of energy to the philosophical primary matter, that of which everything is made. However naive this point of view may (or may not) appear, it is interesting in that it may help to clarify and order things at a conceptual level. 
\par
From the point of view of physics, Heisenberg has no doubt perceived that energy lies at the most fundamental level of all `things', as being constitutive of all that \emph{is}. This interpretation, though, ignores other important quantum facts$\,$:
\par
 1- The mass/energy equivalence, with some anteriority of the latter over the former, has not discarded the concept of mass as the crucial and recent enough \emph{Higgs boson} discovery testifies. 
 \par
 2- The virtual photon just alluded to above, $\gamma^\star(k)$, is not a pure energy, but an energy expressed in the photonic field. Energy is really distinct from what receives energy. This is the reason why energy is communicable, passing from some way of being received in a given quantum field, to another way of being received in another field. \par
 3- Quantum fields are in potency of measurable energetic excitations, their associated elementary particles to begin with. But quantum fields cannot be switched off by depleting their energy content down to zero \cite{A3}: And so, they are not simply made out of energy.
\par\medskip

 \noindent Moreover, the assertion that energy is the quantum fields’ substrate stumbles upon the fact that heat also is energy. Indeed, in $QFT$s, a heat temperature (that is an energy through the relation $E=k_BT$, where $k_B$ is the Boltzmann constant), shows up as an extrinsic factor~${}^{\footnotemark[9]}$\footnotetext[9]{In $QFT$s, in effect, a heat temperature is implemented as formal `$KMS$ boundary conditions' to be satisfied by field configurations (or fields' operators, according to the quantisation procedure being followed)~\cite{Landsman}.}. If energy was the matter of quantum fields, $QFT$s shouldn’t be able to cast heat aside, that is, a valuable part of its matter, without compromising its effectiveness. 
\par
All in all, these reasons point toward the necessary distinction between energy and what is irreducible to energy at the fundamental level of the physical reality:
Energy is the key factor that drives propagations in the fields (Equations (\ref{photon})-(\ref{fermionic2})), quantisation rules (Equations (\ref{quant1}), (\ref{quant2})) and interaction vertices (Equations (\ref{emint})-(\ref{em})) into physically understandable aspects. But energy does not equate to any of these aspects. It is rather what comes to actualise all these formal and operative properties that are proper to the fields’ natures.  A real distinction must therefore be acknowledged between energy and the formal properties.
Energy is what realises differences among the various quantum fields, in that it makes interactions real and brings about actual differentiated propagations ({\textit{i.e}}, elementary particles), which compare to the lack of these interactions and propagations in the same way as \emph{act} compares to \emph{potency}. 
\par
\par\noindent
There is, then, no pure energy \textit{per se} at the level of relativistic quantum physics, because what energy designates indeed is an activity,
\par
 1) \emph{in} a quantum field,
 \par
  2) which, even if deprived of energy, can still \emph{be} ${}^{\footnotemark[9]}$\footnotetext[9]{ It is remarquable (but not by chance \cite{Ax}) that the mathematical formalism is able to testify of this. For systems of interacting quantised fields without derivative couplings ($QED$, for example), energy is an additive property. It follows from the positive character of energy that in the vacuum state of such systems, the various fields fluctuate into each others, each at zero energy: They \emph{exist} at energy zero.}.

\subsubsection{Not a thing }
In a second place, energy is not a thing, or a substance \cite{Hecht}, but something that a thing, more precisely an entity, possesses. 

\par\medskip
 The virtual photon just evoked, which can achieve energy transfers and transformations between elementary particles, provides an illustration of this point. The virtual photon $\gamma^\star(K)$ is an entity sharing with real photons (as physics commonly defines them) some of the determinations which are characteristic of real photons. In the metaphysical wording of \cite{A3} we call these characteristic determinations the \emph{photonic form}.  Accordingly, it is not a pure energy in itself, an energy totally deprived of form. 
 \par
 Besides, there is no reservoir of a pure energy, possibly infinite, as was unduly suggested by the famous \emph{Casimir effect} \cite{A3, Jaffe}; and since then, as it keeps being almost always assumed in the scientific literature. Energy is always \emph{the} energy of something, and more precisely, the energy received in some entity or possessed by some entity, that is, quantum fields. Moreover, the energies received or possessed in quantum fields are not necessarily actualised in determinate modes, like elementary particles \cite{ Hegerfeldt} (or other space-time manifestations~\cite{Georgi}).

\subsection{Energy is not primary matter}

As posited by Heisenberg, this peculiarity of energy, such as briefly recalled above, suggests that energy could be what everything is made of. In short, energy would be this primary matter long conceived by philosophers and composing in all of the material things of the Universe. This point is non trivial at any standpoint, physical or philosophical. In the scientific literature proposals have been made to identify this primary matter out of quantum considerations \cite{dEspagnat}. In \cite{A3}, the identification of primary matter to energy is refuted and it may be appropriate here to recall and specify some further arguments in this direction.

\subsubsection{If energy were primary matter}

In this subsection, things will be taken from a somewhat different point of view by examining the consequences which would follow from an identification of energy with primary matter.
\par\medskip

$\triangleright$ {\textit{From the quantum fields' immutability}}.
\par
Quantum fields reveal themselves as \emph{media} of elementary particles' propagations and of their mutual interactions, and this peculiarity points to their very nature of fields. Now, one thing is for an entity to be in the field and a different thing is what the field itself is made of. Accordingly, it is not because a field can encapsulate energy that it is itself made of energy. Besides, if energy was primary matter, then elementary quantised fields would be made of energy, and in particular, movements would be movements {\emph{of}} the fields themselves, rather than movements \emph{in} the fields \cite{A3, Zee}.
\par\medskip
$\triangleright$ {\textit{From the various levels of quantum realities' actualisations.}}
\par
In the general case, quantum realities display various levels of actualisations, all of them being part of a differentiated physical reality \cite{A2}. Were elementary quantum fields be made of energy, the distinction real/actual versus real/potential would become indistinguishable. An instance of the former is given by the actual propagation of a real photon in the photonic field, as a medium, a `mattress' in which the photon propagates \cite{Zee}. An example of the latter is provided by the quantum fluctuations of the elementary fields' vacuum state, at nul energy, $P_\mu|0>=0$. Though a physical reality, in effect, these fluctuations are not actual realities but pure potential realities ${}^{\footnotemark[10]}$, deprived of any amount of energy.\footnotetext[10]{From \cite{A3}: {\textit{With these vacuum fluctuations it is worth pointing out that one is facing a reality totally deprived of actuality, an exceptional fact in the whole realm of physics which only $QFTs$ have been able to render manifest.}}} If this distinction is not made, supposing fields made of energy, fields become realities endowed with a permanent movement. What they are not \cite{A33}.

\par\medskip
$\triangleright$ {\textit{From the distinction between the quantity of a thing and the `quantity' of an activity in this same thing.}}
\par
This is a philosophical and summarised wording of the following physical consideration. Considering a quantum field expression such as, for instance, that of the photonic field (\ref{photon}), it would seem formally that if the energy ${\vec{k}}$ disappears, then the field $A_\mu(t, {\vec{x}})$ itself disappears, that is, that the total amount of ${\vec{k}}$-energies represents the matter the quantum field $A_\mu(t, {\vec{x}})$ is made of. 
\par
It is not so. These expressions are relative to the energies of the fields’ actualisations, and not to the energy the fields would be made of : It should be clear enough \cite{A3, A2} that the ${\vec{k}}$-energies to be summed upon in (\ref{photon}) refer to the energies of the various photonic actualisations the photonic quantum field is capable of, and not at all to the energy the quantum field would be made of. This crucial observation complies with the keen remark of H.B.G. Casimir that it is impossible to extinguish an elementary quantum field by quenching its energy down to zero; and thus, as stated earlier, an elementary quantum field is not made out of energy. Energy is \emph{not} the matter of quantum fields, and, accordingly, not \emph{the} primary matter in the case of truly elementary quantum fields.
\par\medskip
$\triangleright$ {\textit{From the distinction between primary matter and the matter of corporeal bodies.}}
\par
Eventually, given that there is no reservoir of a would be `pure energy', there is but one possibility for the elementary fields to be made of energy. It is that the elementary fields are compounds of actualised underlying realities, themselves made of other \emph{forms} and \emph{matters}. In this case, elementary fields cease to be elementary and the search for elementarity, together with its associate primary matter, is just postponed one step further.
\subsubsection{On the importance of the differentiation of matter and energy regarding the existence of movement at the most fundamental level}

\par
 Physics has long accustomed us to distinguish between the mobile (e.g. a ball) and its movement (e.g. local), between the matter of the mobile (e.g. a steel ball) and the activity of the same mobile (e.g. rolling), between the mass of the mobile and its energy. These distinctions have been the cornerstone for understanding the laws of motion because they reflect the plurality of principles involved in motion: What is moved is not the same as the power which moves. 
\par
Then came quantum physics: The more physics got closer to matter the more energy it found. Energy and mass appeared intrinsically correlated, hadronic matter revealed as enclosing incredible quantities of energy, and quantum fields showed as the place of innumerable, unceasing activities. It is no surprise then that physicists were spurred to question the fundamental distinctions they used to make. The tendency to identify prime matter with energy came as a natural temptation. The aforementioned reasons should be enough though to resist ideas propelled by imagination and to stick to actual physical results. 
\par\medskip
But another basic point has also to be considered, which is to not undermine the fundamental distinction between what is moved and the power which moves, that is, the philosophical threefold principles on which the notion of movement relies: ` Mobile (= form + matter) + energy'.

In elementary quantum fields now, activities consist of propagations and interactions and are the effects of energy exchanges between fields: If there is motion in a field, it is because the field receives energy, quantum fields displaying their natural ability to encapsulate energy. If there is interaction between fields, it is because the energy that one receives is the energy that another communicates. There is thus a real necessity to distinguish fields (form + matter) from the energy that makes for activities within or between fields. Conflating energy with fields’ prime matter amounts to foreclose any physical \emph{and} philosophical explanation of motion at the elementary scale.
\par
At the elementary scale of quantum fields, in effect, the importance of fields' fluctuations at zero energy is that they clearly identify the mobiles, {\textit{i.e.}}, the fields as given by their forms and their (primary) matter, while energy completes to the relevant number of three, the structural principles of the movement, as recalled above.  
\par

At this fundamental scale, movement must of course be understood according to the nature of the mobiles it affects, that is the elementary quantum fields. So it's about movements \emph{in} the fields, not movements \emph{of} the fields. This distinctive feature taken into account, one can see how the distinction of matter and energy allows one to preserve the notion of movement at the most fundamental level of the physical reality.


\subsection{What should be retained}

$\bullet$ Summing up the preceding subsections, at the deepest level of elementary particles and associated quantum fields, energy and mass are \emph{equivalent} in the very sense of Einstein's relation $E=mc^2$, and effectively, experimental protocols exist allowing to go from $m$ to $E$ and from $E$ to $m$. 
\par
$\bullet$ Now, this does not mean that mass and energy are \emph{identical}. If it turns out that energy accounts for the most part of the visible Universe mass, a tiny but essential part of $1\%$ is apparently supplied by the famous Higgs massive scalar field \cite{Higgs}. 
\par
$\bullet$ There is no pure energy, that is an energy which would be in itself and by itself, totally bereft of any sort of the physical determinations which specify a quantum entity and, {\textit{a fortiori}}, a thing. Energy is always the energy of some thing, were this thing as virtual as a virtual photon $\gamma^\star(k)$, or as potential as a quantum field \cite{A3}. 
\par
$\bullet$ Being in every thing, for the things \emph{to be} and \emph{to be in order to} ${}^{\footnotemark[11]}$, \footnotetext[11]{The zero-energy photon is an individuality of reason but it is not a physical reality. Photons can be measured. They are an actualised and thus measurable aspect of the non-measurable quantised photonic field. But a zero energy photon cannot be measured.}energy seems to be this universal matter out of which everything is made, as some physicists thought, in the same way as, in antiquity, Heraclite thought of the element \emph{fire}. Despite this, though, energy cannot be consistently thought of as the primary matter composing in all of the physical realities of the Universe. As a consequence, Heraclite's position on the subject must be abandoned, however supported it has been by W. Heisenberg himself :
\par
\emph{We may remark at this point that modern physics is in some way extremely near to the
doctrines of Heraclitus. If we replace the word `fire' by the word `energy' we can almost
repeat his statements word for word from our modern point of view \cite{WH}.}
\par
$\bullet$ Energy is neither identical to the primary matter of quantum fields, nor a reality external to the quantum fields. There is no reservoir of a pure energy outside of the quantum fields~${}^{\footnotemark[12]}$. \footnotetext[12]{One may observe that such a reservoir of a pure energy could also render superfluous the law of total energy conservation which physics shows as being carved in the marble over all possible scales.}From a philosophical point of view, moreover, energy stands on the side of act, of which primary matter, by definition, is totally deprived.
\par
An interesting by-product of the distinction between energy and primary matter is that the philosophical notion of motion keeps being relevant to the most elementary scale of physical reality, whereas it would be difficult to maintain it otherwise.
\par
$\bullet$ Eventually, energy is much more in the binding energies of elementary particles into the hadronic matter composing the atoms, and ultimately the substances, than in the masses of the involved elementary particles; so that an atomistic view of the Universe cannot be maintained either. 

\section{At larger scales. }
At larger scales there is no need to recall the many facets of energy that we are long used to and which are essentially summarised in the penultimate quote of Heisenberg that `Energy can be changed into motion, into heat, into light and into tension. Energy may be called the fundamental cause for all change in the world'.  At larger scales, though, things differ in some respects from the fundamental scale aspects. 

\subsection{Mass and energy}
The fundamental scale, that of the elementary constituents of the World, has displayed the equivalence of  mass and energy, the realm of $E=mc^2$. Higgs physics is only 12 years old and still underway, but if its crucial point is preserved by whatever new progress ${}^{\footnotemark[13]}$, \footnotetext[13]{The Higgs field is \emph{scalar} and this only, is an issue in itself: Do scalar fields really exist as elementary ones? Some authors have recently suggested that the Higgs field, of mass $125\, GeV$, would enjoy a further resonance at some $700\, GeV$, which would seriously question its supposed elementarity \cite{Higgss}.}one can learn of it that mass would enjoy an origin other than energy. In other words, if the Higgs field is at the origin of all of the elementary particles' masses, then, mass cannot be indefinitely reduced to energy :\par
The conversion of the Higgs particle and its mass into an energy of $E=m_Hc^2$ radiated by gluons, remains possible, and likewise, the reverse process known as \emph{gluon fusion} into a massive Higgs particle, of mass $m_H=E/c^2$, as pictured in Fig.5.
 \begin{figure}[!htb]
   \includegraphics[scale=0.3]{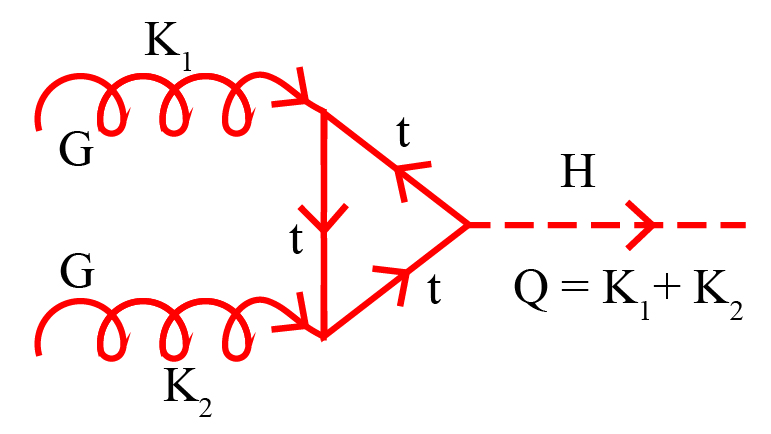}
   \caption{The fusion of 2 gluons, through a top quark (t) triangle, emitting a Higgs particle $H$ of total 4-momentum $Q=K_1+K_2$. Conservation of total 4-momenta at each vertex of interaction. Contrary to the case of Fig.4, gluons in this case can be real, {\textit{i.e.}}, $K_i^2=0$.}
   \label{fig 5}
\end{figure}  But in the Standard Model of elementary particles, for the weak bosons and the quarks to acquire a mass, an irreducible massive and fundamental Higgs scalar field is mandatory. In this sense, some possible form of irreducibility of mass ${}^{\footnotemark[14]}$, \footnotetext[14]{Irreducibility.. \emph{to energy} only!  In effect, the Higgs mechanism proposes an elucidation about the generation of the elementary particles' masses;  could mass in this way, still be considered as a primary quality of matter?. Not yet decided, we think.}long taken as a proven fact in classical physics, would show up at the level of elementary fields themselves. As pointed out earlier,  
\par
\emph{Newtonian mechanics posited mass as a primary quality of matter, incapable of further elucidation. We now see Newtonian mass as an emergent property \cite{Wilczek}.}
\par\medskip
The realm of $E=mc^2$ is not unlimited, though. The fact that the Einstein famous relation is a truth on which a lot of fundamental aspects hinges, does not make of it an \emph{operational} truth at every level of the physical reality. This statement is related to a meaningful feature of this physical reality:
\par
If at the elementary scale $E$, as $mc^2$, and $m$, as $E/c^2$, can be realised both, now no protocol ever would transform a cat into energy ... were it the enigmatic Schr\"odinger's cat itself! 
\par
Such unrealistic an attempt would proceed from the same conceptual abuse, which consists in an unlimited reductionism approach, reducing proven substantial realities down to the quantum behaviours of their constituents \cite{A1}; while, again, Wisdom and experience would rather suggest that \emph{More is different} \cite{More}.
\subsubsection{Mass' autonomy}
What happens in deeds is that at larger scales, mass acquires its autonomy with respect to energy. 
\par
At our human scale for example, mass has clearly acquired its autonomy with respect to energy, as classical physics amply testify, where one passes from masses weighted in energy units, $eV, MeV, GeV, $ to masses weighted in grams, kilograms and tons. 
\par
As announced in footnote [5], the famous Einstein's equation $E=mc^2$ is not \emph{relativistically invariant}, but holds true in the particular inertial reference frame in which the mass $m$ is at rest. The invariant equation, valid in any inertial reference frame, reads instead,
\begin{equation}\label{invariant}E^{\,2}=m^{\,2}c^{\,4}+{\vec{p}}^{\ 2}c^{\,2}\end{equation}where ${\vec{p}}$ is the body's \emph{$3$-momentum}, or \emph{impulsion} or \emph{quantity of movement}. In the jargon of physicists, (\ref{invariant}) represents a \emph{covariant} expression, while $E=mc^2$ doesn't. This covariant equation, which applies to both classical and quantum physics, is very interesting because in it, the mass $m$ is a relativistic invariant, contrarily to the quantity of movement or impulsion, ${\vec{p}}$, a body possesses and which can be subject to continuous variations.
\par
This shows what happens when bodies are constituted. Their masses can become part of their definitions, as they are attached to what individualises them as substances. Masses relate to what make up bodies in their consistency as bodies, and no longer to what happens to them (see below).
\par
Up to a few formal definitions given below, it is remarquable that this state of affairs possesses a clear enough philosophical translation which helps to understand the nature of what we talk about. 

\subsubsection{ The intertwining of energy and mass : an account of stable substances}

By highlighting the various aspects of the relationship between energy and mass, physics provides a better understanding of the constitution and behaviour of bodies on large scales. What characterises these bodies is that they enjoy some relative stability through their movements. A steel ball is able to roll along while keeping on being a steel ball. For the sake of philosophical accuracy, such bodies can be called substances, while their capacity to encapsulate an energy which is subject to autonomous variations can be called accidental.
\par\noindent
Now, apart from their accidental energy, we know that these substances are not deprived of energy in themselves. But this inner energy is possessed ‘in principle only’ (not in an operational way ${}^{\footnotemark[15]}$\footnotetext[15]{Keeping in mind that no experimental protocol would ever transform a cat into an atomic bomb, as quoted above.}), their $mc^2$ is fixed because mass is a relativistic invariant. And in this sense, the mass, or proper energy a stable substance possesses in itself is part of the substance’s definition: It is part of the definition of a steel ball that it has a mass, or proper energy.
\par\noindent
Besides, a substance has a capacity to encapsulate energy, attached to the ${\vec{p}}^{\ 2}c^{\,2}$-piece of (\ref{invariant}), which is obviously subject to independent variations, and this capacity is therefore \emph{accidental}.
\par\noindent
Therefore, energy in a substance is dual, which means that it has two distinct levels of actuality or activity. The substance’s mass, or proper energy, makes for a first act, and the accidental energy, subject to variation, makes for a second act.
\par\noindent
The situation is different at the quantum level of the physical reality. The language of substance and accident is irrelevant (quantum fields are definitely not substances \cite{A3}). But the language of first and second act is still on point. Quantum fields are indeed able to encapsulate energy subject to variation. An energy which activates the field can therefore be construed as the field’s second act. But the very same energy is also what constitutes the first act of more elaborate/actualised physical realities. This deserves further illustration.

\par
 In the first place, quark fields \emph{are}, and this defines their \emph{first act}, {\textit{i.e.}}, their act of being. Then, finite energy excitations, {\textit{i.e.}}, elementary quark particles, ebb out of them. This defines the quark fields' \emph{second act}, which, as evoked above, is also their proper mode of actualisation \cite{A3}. From these elementary quark particles, the Hadronic matter is generated subsequently, a proton for instance ${}^{\footnotemark[16]}$.\footnotetext[16]{This is also what avoids being deceived by naive representations in which elementary quark particles are pictured as small balls inside a proton. One should keep in mind that, in a quantum field, the finite energy excitations known and reported as the associated elementary particles, are defined through well defined experimental/theoretical conditions \cite{A3}, which are absolutely not met in the inside of a proton.}
 
 \par
In this sequence, the elementary quark particles, which result from the quark field’s activity (its second act), serve as matter to constitute the actual proton (the proton’s first act). Energy is then used to constitute mass as we saw above. This applies to any elementary particle stemming out of its corresponding quantum field. This process is caused through interactions. For instance, in a Drell-Yan processus, the virtual photon $\gamma^\star(K)$, which results from the activity in the quantum photonic field (this field's second act), acts upon the quantum leptonic field, and communicates its energy so that from the leptonic field a lepton pair $l^+,l^-$ is created (an $e^+e^- $ pair for example, as pictured in Fig.3).
\par
This very sort of a sequence, now, where the second act of a physical reality (the field) is constitutive of the first act of another physical reality (its associated particle), is precisely what is definitely excluded from the larger scale physics, where the corporeal realities second acts are essentially reduced to movements in real spacetime, as long translated into the corpus of classical physics. 

\par\medskip
To put things in a coarse, not rigorous but maybe more intuitive way, one could say that if energy can be viewed as a certain measure of the fields' actualisations, mass could be seen as a measure of the tendency of some entities to become the matter of more elaborate forms, corresponding to real, massive and localised substances; what we are used to call `things' in ordinary language.
\section{What is energy ?}
\subsection{Nobody knows}
Most often, simple questions are not the easiest ones to be answered. Intuitively, one might think that energy would find its explanation along with the uncovering of the smallest components of physical reality. The more we understand what things are made of, the more we should expect to understand what energy is. For the most reductionist minds, this should even lead to explaining the origin of the universe.
\par
Whereof the inclination to think of energy as the prime matter from which everything comes. Whereof also the more popular proposal which amounts to conceive the whole Universe as generated by an energetic enough quantum fluctuation of the vacuum: A nonsense in view of the current and previous analyses \cite{A3}.
\par
However, as we hope to have made clear enough through these analyses, what appears instead at the smallest scale is that energy is an irreducible component of physical reality, which doesn’t emerge nor decay but only transfers through interactions, and which doesn’t exist in itself but only as the activity or actuality of things, accounting for the movements through which things are constituted, are acted upon or act upon.

\par\medskip
It turns out then that energy is tightly related to the \emph{existence} of things. It is the condition \emph{sine qua non}  for the things \emph{to be} and \emph{to be in order to..}, the physical realities corresponding to quantum fields and their associated elementary particles, to begin with. 
\par\medskip\noindent
This may be the more accurately perceived by focusing on an aspect of quantised fields which greatly amazed the dutch physicist H.B.G. Casimir, a point which has been evoked already in these lines. Contrarily to classical fields, in effect, quantum fields cannot be switched off \cite{ Casimir}. This striking peculiarity turns out to be perfectly coherent with the understanding of the quantum reality such as proposed so far \cite{A1,A2,A3}. 
\par
 Things can be viewed as follows. A classical field integrates a very large number of more elementary components, as so-called {\emph{coherent states} display in an explicit manner at the formal level. For these states, which are quasi-classical states, components are like a matter entering their constitution. Then, in principle, {\textit{i.e.,}} whatever the experimental way to do it, it is possible to act upon these components to bring the coherent states down to zero. In an analogous way, here comes the crucial difference with an elementary quantum field. Elementary quantum fields are the very first physical determinations of the material World, and they do not proceed from a composition with a pre-existing matter, whatever would be the degree of actualisation of this would-be pre-existing matter ${}^{\footnotemark[17]}$\footnotetext[17]{The matter of quantum fields is the \emph{primary matter} such as identified in \cite{A3}. It is universal and unique in that it is a reality totally deprived of actuality and made manifest only through the vacuum fluctuations of elementary quantum (and quantised!) fields.}. They show up, exactly as they are, and there is no matter on which to act to switch them off. It is not possible to create an elementary quantum field, and likewise, to make it vanish. 
 \par
 Intuitively, this conclusion is certainly not completely astounding, but from a philosophical point of view, it is not innocuous either.
 
 \subsection{What can be stated}
Thus, even deprived of any amount of energy, something of the elementary quantum fields persists, their vocation to interact with other fields included, as is made clearly manifest by considering their vacuum fluctuations. However, the crucial point, reminded in footnote [9], is that these vacuum fluctuations remain a purely potential reality, deprived of any spacetime anchorage: There is no possible spacetime description of the vacuum quantum fluctuations.
Fully potential, bereft of actuality and nevertheless a part of the physical reality, such are the zero energy quantum fluctuations.  Quantum field theories only have been able to display this state of affairs \cite{A3} and this is most relevant when it comes to understand what energy~is.
\subsubsection{A physical characterisation and two philosophical indirect definitions}
{\textbf{-}} From the physical apprehension of the zero energy vacuum fluctuations, in effect, energy shows up as the component which allows the elementary corporeal realities ({\textit{i.e.}}, the excitations in the elementary quantum fields, formally described in $QFTs$ by the quantised fields' correlation functions) to enter existence in the very sense of entering spacetime actuality through measurability. 
\par
{\textbf{-}} The physical characterisation just proposed \emph{must} also be complemented with the following other caracterisation: Energy is the actual quantification of a natural power to interact, that is, `act upon' or `receive from'. That those powers exist as capacities or dispositions, and that they are natural to quantum fields, appears with the interaction diagrams that $QFT$s have identified and inventoried ${}^{\footnotemark[18]}$\footnotetext[18]{One may note that this is in line with contemporary thinking on causal dispositions developed from an analytical perspective~\cite{Tiercelin}.}.
\par\noindent
It is interesting to observe that these two characterisations of energy relate precisely to the two aspects of the elementary constitutive elements of the material universe, which \emph{are} and\emph{ are in order to} ; and which energy projects, both, into the physical existence in spacetime in close correlation with measurability.
\par\medskip\noindent
Now, it must be clearly stated that energy shows up as a primary notion. Contrary to mass, which, at least up to the still enigmatic Higgs scalar field, is reducible to energy, there is no antecedent notions out of which to derive the notion of energy. This explains in the most cogent fashion that `nobody knows what the actual energy is \cite{Web}'.
\par\medskip\noindent
 In order to correctly understand this physical apprehension of energy, two indirect definitions of it can also be given supporting it with a more metaphysical expression,
\par
 {\textbf{--}} First, with respect to the primary matter: Energy isn’t primary matter and is \emph{opposed} to it, as \emph{activity} is opposed to pure potency.
 \par
 {\textbf{--}} Second, with respect to the formal properties: Energy is the act of a formal property (for instance, the proper actualised values of the capacity an elementary particle has to bear definite impulsions). The formal property is itself intrinsically ordered to the act made possible by the energy. And energy relates to the formal property in the same way as the act relates to potency.
\subsubsection{Are energies of the Universe finite?}


Based on a mix of philosophical and physical arguments, a few more considerations can be ventured concerning the possible, if not plausible, finite character of the energies met in the Universe. Of course, it is notoriously difficult, if not even naive, to decide on the finite or infinite character of the energy of the Universe as a whole, be it, first of all, because the `whole Universe' as a notion is far from being circumscribed.
\par

It remains that, what our current understanding of energy in quantum realities shows, namely that energy is always received, and that it is limited by this reception in something, is that:
\par
\noindent
If one makes the energy of the Universe tend towards infinity, the result should not manifest itself in realities of infinite energy, but rather in a proliferation (tending to infinity?), of realities endowed with a finite energy. In this sense, whether the energy of the Universe is finite or not, it is possible to venture that the universe is filled with a finite and conserved energy.

\emph{Measurability} comes about as a consequence of \emph{actualisation}, in that the reality is \emph{finite}: The act is received by a potency ${}^{\footnotemark[19]}$\footnotetext[19]{Potency is not to be conceived as a container but as a correlative or a co-principle, in the sense that neither exists, the act and the potency, in a separate state.} which provides it with some individualisation, in the same time as it individualises it in a {finite} actuality. The \emph{necessity of finiteness} can be recognised in the fact that energy is met within formal properties only, and this constitutes for energy an intrinsic limitation. 
\par\noindent
In particular, this necessity can also be appreciated with respect to \emph{intelligibility}. What could possibly mean a formal property endowed with an \emph{infinite} energy? Which sort of a knowledge could possibly be acquired from it? ${}^{\footnotemark[20]}$\footnotetext[20]{Infinite means unbounded from above and thus undefined. Such a formal property would therefore be unable to be the member of any comparison to other existing measurable physical realities: Deprived of any relation to the apprehensible physical world, no intelligibility of it could be acquired.}
\par
One could object that the self energy calculations performed in perturbative expansions of $QFT$s are plagued with divergences and yield infinite results. Are thus elementary particles's self energies infinite? From 1930 on, this problem, tamed technically with the renormalisation theory,  has puzzled physicists for several decades regarding its interpretation, and it is not so long ago that the magics of the most efficient \emph{renormalisation theory} could eventually be uncovered.
\par
Not seeping here into numerous and very technical details, some crucial points are in order for our current concern. First, it must certainly be acknowledged that a modern understanding of $QFT$s amounts to consider them as \emph{low energy effective theories}. This means that the infinities induced by the very high energetic quantic fluctuations are not physically relevant. $QFTs$ are valid up to an upper energy limit provided by a cut-off parameter $\Lambda$, say, and the problem of infinities is simply repealed:
\par
{\textit{Who is afraid of infinities? Not I, I just cut them off \cite{Zee}}}
\par\noindent Now, while this interpretation can make full practical sense, in some cases at least, it may be not so fundamental as one thinks it is. This is because a much deeper and revealing fact is that the divergences met in calculating self energies cannot be associated to any physical reality.  In effect, as self energies are expressed in terms of \emph{measurable} quantities, then all of the intermediate infinities met in the calculations simply disappear from the final results ({\textit{i.e.}}, the dependences on the cut-off parameter $\Lambda$) which, moreover, display a remarquable predictive power extending to a very large span of measurable properties. 
\par
To sum up, $QFT$s which inherently call for renormalisation in order to make sense, do not testify to the existence of any physical reality endowed with an infinite energy. Quite on the contrary.

\noindent From another physical point of view, that of large distance scale physics, if some formal property was associated to an infinite energy, so far as is known from gravitation theory, this formal property's first act would weigh so much that it would induce the accretion to itself of the entire Universe. So far, there is no indication of anything like this. 
\section{Conclusion}
\emph{Energy is a very subtle concept. It is very very difficult to get it right}, as is announced in the Introduction, recalling the point of view R.P. Feynman expressed on the matter \cite{Feynman}. However, a deep theoretical physics analysis relayed by some philosophical clarifications allows one to conclude to the following short series of statements.
\par\medskip

Energy is not a thing, but a necessary component of the things. 
\par
It is not a property of things, but an actuality of their property.
\par
It does not exist outside of things, but within the things themselves.
 \par
 It is not a matter, or \emph{the} matter, but it is what pulls matter out of its potentiality. 
 \par
Energy is what makes up the actuality of the material World, from the very first actualisations in quantum fields, up to the constituted stable substances of spacetime.
 \par
It is through all of these actualisations that the Universe lends itself to measurability and intelligibility in a second step. 
 \par
 Energy is a \emph{datum} of \emph{Nature}, on the same footing as primary matter and the finite number of elementary fields are \emph{data} of Nature.
 \par
Neither does energy appear nor does it disappear, but it communicates itself in such a way that studying the actualisations, actions and movements is our entrance gate into the knowledge of the material Universe fundamental principles at both technical and philosophical levels.
  \par
 The Universe is full of a finite and conserved energy in the same way as it is made out of a primary matter which is pure potency, and of a finite number of elementary quantum fields. Even when deprived of any energy, the latter are still endowed with  a sort of `ghost-like' manifestation, their vacuum fluctuations, attached to sets of well defined physical properties, however devoid of actual realisations.
 \par
Structuring our material Universe one must therefore distinguish: 1- A primary matter \cite{A3} (\emph{primary matter}: In potency to everything and therefore unmeasurable due to lack of any determination and of any act), 2- the formal properties, including their natural operative power to interact (\emph{primary matter + formal properties}, {\textit{i.e.}}, the case of elementary quantum fields: In potency to everything that formal properties can produce, but not yet measurable for lack of act), 3- the acts,  which are the acts of formal properties, acting on or acted upon, and as such, are these formal properties’ actual quantified energies (\emph{prime matter + formal properties + actuality}: Becomes measurable because the act is the actualising principle of a material form, \textit{i.e.}, the receiving principle that limits and individualises. This is the case of elementary particles and of their subsequent compounds). 
\par
The study of energy shows that the building blocks of the material Universe are not elementary quantum fields alone, but quantum fields together with their interaction diagrams, and it also pinpoints the most crucial aspect of energy which, through differentiated physical actualisations, relates tightly to the Universe's existence, measurability and intelligibility.

\end{document}